\documentclass[onecolumn,prl,english,showpacs]{revtex4-1}
\usepackage[latin9]{inputenc}
\pdfoutput=1
\usepackage{amstext,bm}
\usepackage{graphicx}
\usepackage{xcolor}
\usepackage{color}
\usepackage{amsmath}
\usepackage{babel}
\usepackage[titletoc]{appendix}
\usepackage{hyperref}
\hypersetup{
     colorlinks = true,
     citecolor  = magenta,  % cite
     linkcolor  = magenta % ref
}
\usepackage{float}
\newcommand{\red}[1]{\textcolor{red}{#1}}

\begin{document}

\title{From Nodal Chain Semimetal To Weyl Semimetal in HfC}

\author{Rui Yu$^{1}$}
\email{yurui1983@foxmail.com}
\author{Quansheng Wu$^{2}$}
\author{Zhong Fang$^{3,4}$}
\author{Hongming Weng$^{3,4}$}
\email{hmweng@iphy.ac.cn}

% \selectlanguage{english}%

\affiliation{ $^{1}$
School of Physics and Technology, Wuhan University, Wuhan 430072, China}
\affiliation{$^{2}$
Theoretical Physics and Station Q Zurich, ETH Zurich, 8093 Zurich, Switzerland}
\affiliation{$^{3}$
Beijing National Laboratory for Condensed Matter Physics, and Institute of
Physics,Chinese Academy of Sciences, Beijing 100190, China}
\affiliation{$^{4}$
Collaborative Innovation Center of Quantum Matter, Beijing 100190, China}

\begin{abstract}
Based on first-principles calculations and effective model analysis, we propose that
the WC-type HfC, in the absence of spin-orbit coupling (SOC), can host three-dimensional
(3D) nodal chain semimetal state.
Distinguished from the previous material IrF$_4$
\cite{nodalchain_Nat_2016}, the nodal chain here is protected by mirror
reflection symmetries of a simple space group, while in IrF$_4$
the nonsymmorphic space group with glide plane is a necessity.
{Moreover, in the presence of SOC, the nodal chain in WC-type HfC evolves into Weyl points.
In the Brillouin zone, a total of 30 pairs of Weyl points in three types are obtained through
the first-principles calculations. }
Besides, the surface states and the pattern of the surface Fermi arcs connecting these Weyl points are studied, which may be measured by future experiments.
\end{abstract}

% \date{\today}
\maketitle
\noindent\textit{Introduction. ---}
Materials with novel Fermi surface have stimulated significant interest in the field of
solid state physics\cite{rev_sm_wenghm,nodalchain_Nat_2016,BAB_6Degen_nodalP_Science_2016,Weng_3Degen_nodal_PRB_2016,WQS_3Degen_nodal_PRX_2016}.
The symmetries of crystals give the constraints to the band structure of quasiparticles.
This may lead to the quasiparticles near the Fermi energy behave like Dirac fermions
or Weyl fermions that are first come up within high-energy physics,
or some other novel quasiparticles that have no high-energy correspondence.
For example, topological nontrivial semimetals can host
three-
\cite{Weng_3Degen_nodal_PRB_2016,WQS_3Degen_nodal_PRX_2016,WHM_3Dengen_DSM_PRB_2016,chang_triple_2016},
four-
\cite{WZJ_DSM_PRB_2012,WZJ_DSM_PRB_2013},
six-
\cite{BAB_6Degen_nodalP_Science_2016}, or
eightfold
~\cite{Kane_8Degen_DSM_PRL_2016,BAB_6Degen_nodalP_Science_2016}
degeneracy band-crossing points which are protected by special crystalline symmetries
on high symmetrical crystal momentum points or along high symmetry lines.
The twofold degeneracy bands' crossing points, namely,
Weyl points
~\cite{WanXianGang_WSM_PRL_2011,
XuGang_WSM_PRL_2011,
Weng_WSM_PRX_2015,Huang:2015ic,Xu:2015jx,Lv:2015fj,
Weyl_xu_discovery_2015,
BAB_WSMII_Nat_2015}
are protected by the lattice translational symmetry without other additional crystalline symmetries.
Particularly, in a certain case, in the nodal line
~\cite{GP_Mikitik_NL_PRB_2006,Burkov_NL_PRB_2011,Vivek_NLS_PRB_2014}
and nodal chain semimetals
~\cite{nodalchain_Nat_2016}, the bands' crossing points are not discretely located
in the Brillouin zone (BZ), but they form closed loops.
{So far, three types of nodal line semimetals have been proposed.
Type A is with mirror reflection symmetry~\cite{Schnyder_mirrorsymm_PRB_2014,Hasan_NLS_NC_2016,Be_PRL_XQChen,rev_SM_fangcheng,CaAgX,TlTaSe2},
type B is with the coexistence of time-reversal symmetry and space inversion symmetry~\cite{Weng_NLS_PRB_2015,Kane_NLS_PRL_2015,YuRui_NLS_PRL_2015,Ca3P2,Ca3P2_PhysRevB,Fuliang_2015arXiv,FangChen_PRB_2015,zhao_BP_2016,WangJian-Tao_PRL,Murakami_2017,ZhangFan,DuanWenhui,ChenXing-Qiu_PRL,wanxiangang_2017,xu_CaP3_2017},
while type C is with a nonsymmorphic space group with glide plane or screw axes symmetries~\cite{FangChen_PRB_2015,nodalchain_Nat_2016,LiangQiFeng_PRB}.
The novel band structures lead to exotic properties, including the surface Fermi arc states
~\cite{WanXianGang_WSM_PRL_2011,QXL_WSM_2013},
chiral anomaly in Dirac and Weyl semimetal states
~\cite{WSM_PRB_2013,QXL_WSM_2013,WSM_Exp_PRX_2015},
and drum-head-like surface states proposed in nodal line and nodal chain semimetals
~\cite{Heikkila_2011JETP1,Weng_NLS_PRB_2015,Kane_NLS_PRL_2015}.
}

In the present Letter we propose that the WC-type HfC can host a three-dimensional nodal chain in the BZ
when spin-orbit coupling is ignored.
The nodal chain is composed of nodal lines, which belong to the \red{type A} nodal line structure. These nodal lines are protected by the mirror reflection symmetry in the
$k_z=0$ and $k_y=0$ planes.
This is remarkably distinguished from the nodal chain known in IrF$_4$
~\cite{nodalchain_Nat_2016},
which is protected by the glide plane in the nonsymmorphic space group.
The proposed designing scheme of nodal chain structure in HfC
may be useful for designing nodal chain structure in some artificial systems, photonic and phononic crystal systems.

Similar to TaAs and ZrTe~\cite{Weng_WSM_PRX_2015,Weng_3Degen_nodal_PRB_2016},
in the presence of SOC, HfC can open a gap along the nodal chain,
which leads to three types of Weyl node pairs off of the mirror planes.
Because of the time-reversal and crystalline symmetries, there are 30 pairs of Weyl nodes in total. Here, the surface states and the patterns of the surface Fermi arcs connecting the Weyl points with different Fermi energy are studied.

\begin{figure}[t]
\begin{centering}
\includegraphics[width=0.5\columnwidth]{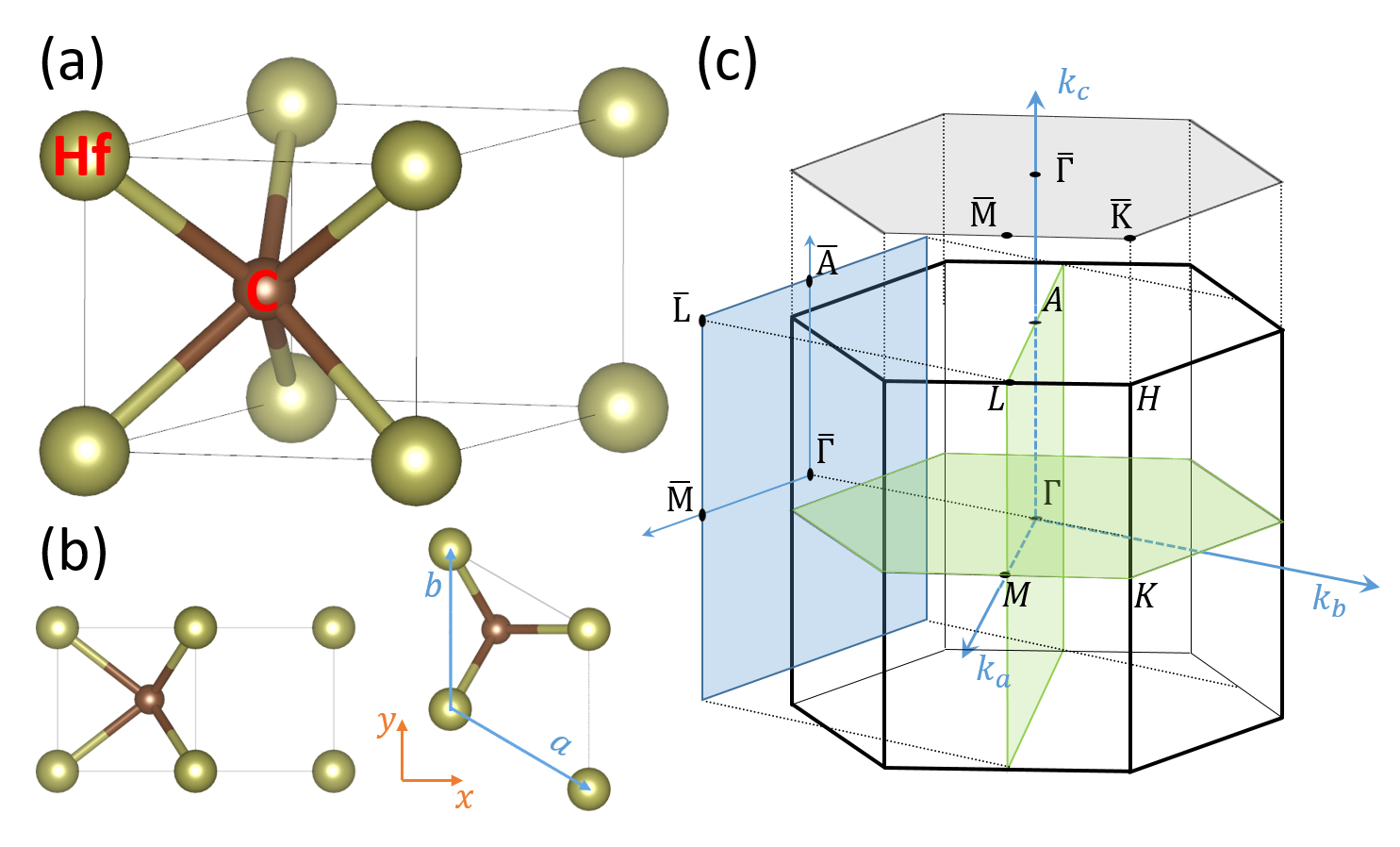}
\end{centering}
\caption{\label{fig:fig1_crystal}
(Color online)
(a) The primitive cell of WC-type HfC and its
(b) top view and side view. Hf and C atoms occupy the $1d$ ($\frac{1}{3}$, $\frac{2}{3}$,
$\frac{1}{2}$) and $1a$ ($0,0,0$) Wyckoff positions.
(c) The bulk Brillouin zone and its projection onto the (010) and (001) direction.
The high symmetrical crystal momenta and $k_z=0$, $k_y=0$ mirror planes are indicated.
}
\end{figure}

\vspace{1mm}
\noindent\textit{Crystal structure of HfC. ---}
The WC-type HfC crystallizes in the hexagonal space group $P\bar{6}m2$ (No. 187).
Hf and C atoms occupy the $1d$ ($\frac{1}{3}$, $\frac{2}{3}$, $\frac{1}{2}$) and $1a$ ($0,0,0$) Wyckoff positions as shown in
Figs. \ref{fig:fig1_crystal} (a,b).
The lattice constants are $a$=$b$=3.267 $\rm\AA$ and $c$=2.942 $\rm\AA$\cite{HfC_crystal}.
All the results discussed in the following are from the calculations with this structure.

% We perform density functional calculations by using the Vienna {\it ab initio} simulation package
% ~\cite{VASP}
% with generalized gradient approximation (GGA)
% ~\cite{PBE}
% and the projector augmented-wave method
% ~\cite{PAW_Blochl:1994uk}.
% SOC is taken into account self-consistently. The surface band
% structures are calculated in a tight-binding scheme that projected from the bulk Bloch
% wave functions  based on the maximally localized Wannier functions (MLWF)
% ~\cite{SGF,Vanderbilt_RMP}.

\begin{figure}[t]
\begin{centering}
\includegraphics[width=0.5\columnwidth]{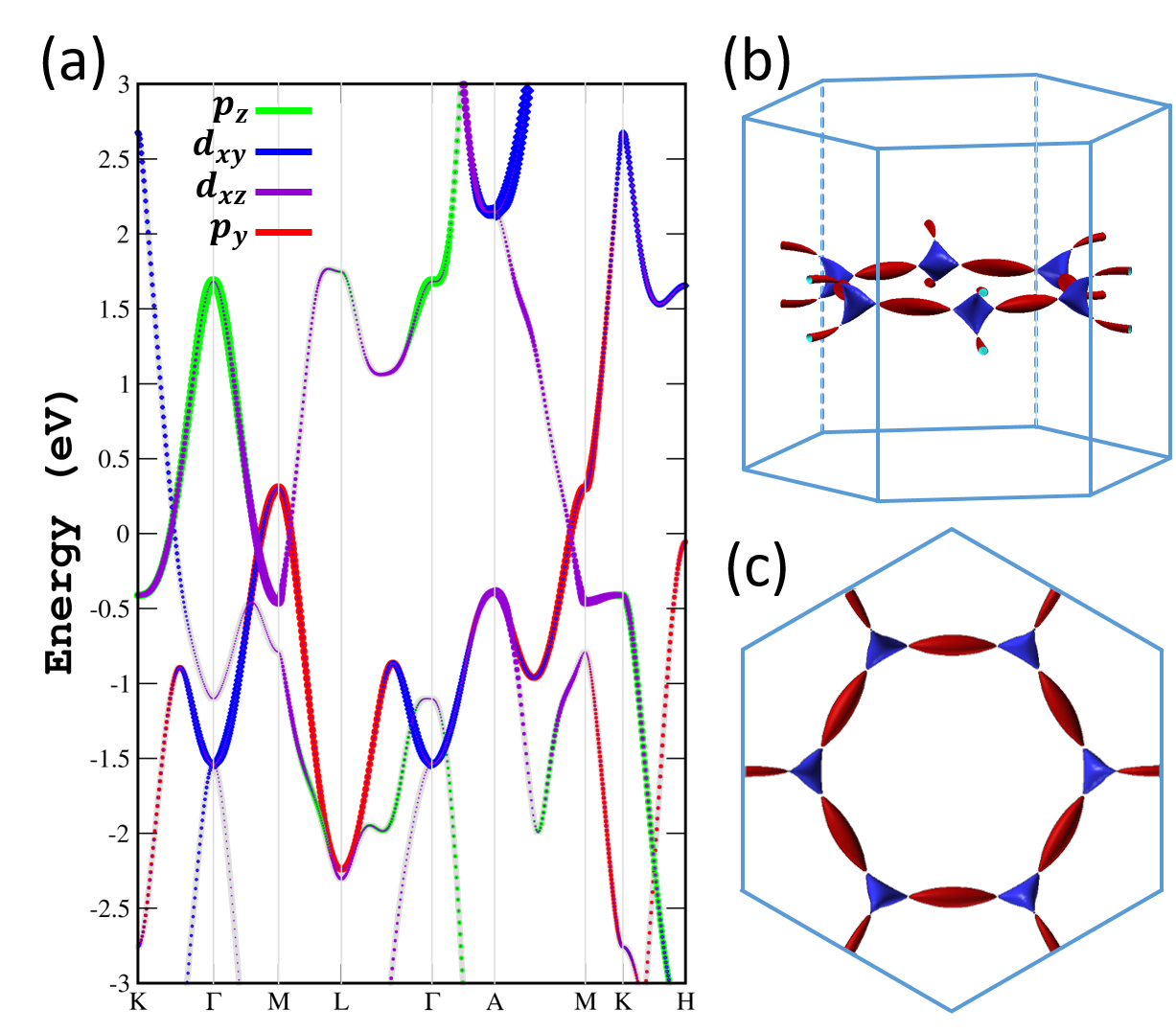}
\par\end{centering}
\protect\caption{\label{fig:fig2_bands}
(Color online)
(a) Band structures of WC-HfC {without SOC}. Two band-crossing structures, one around the $\Gamma$ point
and the other one around the $\rm M$ point, exist near the Fermi energy.
(b) Side view and (c) top view of the Fermi surface of WC-HfC,
which consists of electron pockets (blue) and hole pockets
(red), and forms a lotus-root-like structure.
}
\end{figure}

\begin{figure}[ht]
\begin{centering}
\includegraphics[width=0.5\columnwidth]{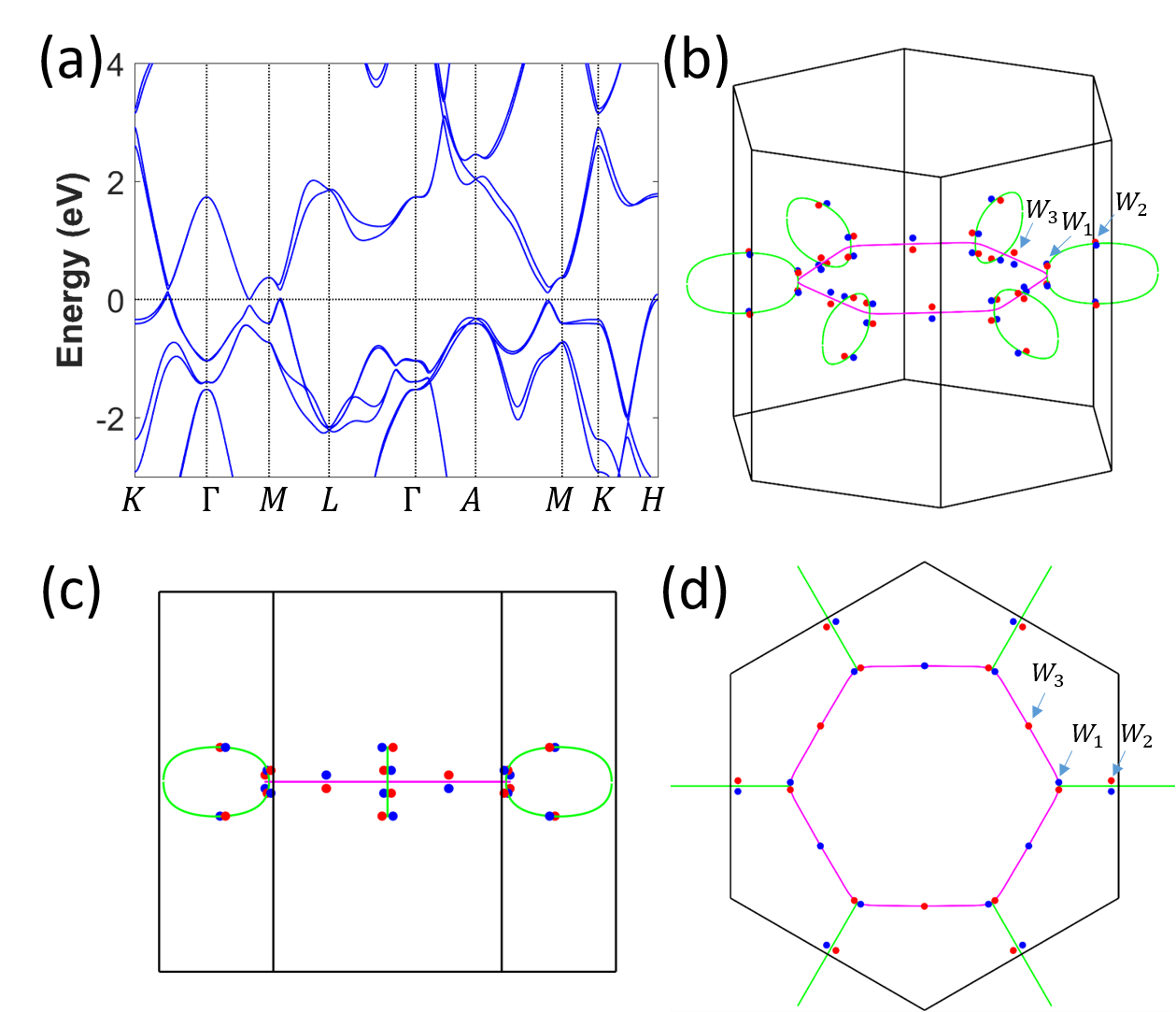}
\par\end{centering}
\protect\caption{\label{fig:fig3_soc}
(Color online)
(a) The band structure of HfC calculated within GGA including spin-orbit coupling (SOC).
(b) 3D view of the nodal chain (in the absence of SOC) and Weyl points (with SOC) in
the BZ. The red and blue points indicate the Weyl point with
chirality $\chi=+1$ and $\chi=-1$, respectively.
(c) Side view from (100) and (d) top view from (001) directions for the nodal chain and Weyl points.
Once the SOC is turned on, the nodal chain are gapped and give rise to 30 pairs of Weyl points off the mirror planes.
}
\end{figure}

\vspace{1mm}
\noindent\textit{Band structure of HfC. ---}
The band structures calculated within GGA are shown in Fig.~\ref{fig:fig2_bands} (a).
It clearly shows that two band-crossing structures exist near the Fermi energy.
One crosses around the $\Gamma$ point and the other one crosses around the $\rm M$ point.
The fat-band structure shows that the two crossing bands around the $\Gamma$ point are dominated by $Hf:d_{xy}$ and $C:p_z$
orbitals for the valence band and conduction band, respectively.
While the band crossings around the $\rm M$ point are dominated by $Hf:d_{xz}$ and $C:p_y$
orbitals for valence band and conduction band, respectively.
{The band-crossing structures near the $\Gamma$ and the $\rm M$ points are confirmed by the nonlocal Heyd-Scuseria-Ernzerhof (HSE06) hybrid functional calculations~\cite{suppMat}.
}
The calculated Fermi surface is shown in Figs.~\ref{fig:fig2_bands} (b-c), where the
surface with blue color comes from the conduction bands and the red part comes from
the valence bands.
Two types of ring structures exist in the Fermi surface. The first
type of rings lie in the $k_z=0$ plane and surround the $\Gamma$ point, the second
type of rings lie in the $k_y=0$ plane and surround the $\rm M$ point.
The lotus-root-like Fermi surface wraps the bands crossing loops, namely, the nodal
chain, as shown in Fig.~\ref{fig:fig3_soc} (b), which lie in the mirror planes that
will be discussed below.

\vspace{1mm}
\noindent\textit{Nodal chain in HfC. ---}
We first prove the existence of a nodal ring encircling the $\Gamma$ point and lying in the
$k_z=0$ plane. The symmetry at the $\Gamma$ point is characterized by the $D_{3h}$ symmetry
group, which includes a $C_3$ rotation symmetry around $z$ axis, $C_2$ rotation symmetry
around $x$ axis, mirror symmetry
$M_y: (x,y,z)\rightarrow (x,-y,z)$ and
$M_z: (x,y,z)\rightarrow (x,y,-z)$, and time-reversal symmetry.
A minimal model for the two crossing bands {around the $\Gamma$ point} can be written as the following two-band
k$\cdot$p Hamiltonian:
\begin{equation}
H_0^{\Gamma}({\bf k})=\sum_{i=x,y,z}{d_i({\bf k})\sigma_i},\label{eq:H_kp}
\end{equation}
where $d_i(\bf k)$ are real functions and the vector ${\bf k}$ is relative to the $\Gamma$
point. We have ignored the term proportional to the identity matrix since it is
irrelevant in the studying the band touching. The time-reversal symmetry operator
is represented by $T=K$, where $K$ is the complex conjugate, in the absence of SOC.
Time-reversal symmetry requires that
\begin{equation}
T H_0^{\Gamma}({\bf k}) T^{-1}=H_0^{\Gamma}(-{\bf k}),\label{eq:THT}
\end{equation}
which places a constraint on $d_i({\bf k})$ so that
$d_y({\bf k})$ is an odd function of $\bf k$, while $d_{x,z}({\bf k})$ are even
functions of ${\bf k}$ and $d_z(\bf k)$ can be generally written as
\begin{equation}
d_z({\bf k})=c_0+c_1 k_x^2+c_2 k_y^2+c_3 k_z^2,\label{eq:dz}
\end{equation}
up to the second order of $\bf k$.
{The parameters $c_i$ can be derived by fitting the dispersions to those of first-principles calculations
\footnote{The fitted parameters are listed as:
$c_0=1.5$ eV,
$c_1=-2.4627$ eV$\cdot \rm \AA ^2$,
$c_2=-2.9791$ eV$\cdot \rm \AA ^2$,
$c_3= 6.0146$ eV$\cdot \rm \AA ^2$,
$a_0= 0.4000$ eV,
$a_1=-3.5749$ eV$\cdot \rm \AA ^2$,
$a_2= 6.1965$ eV$\cdot \rm \AA ^2$, and
$a_3=-5.9582$ eV$\cdot \rm \AA ^2$.}.
The signs of these parameters are the key factors for the existence of nodal rings which can be obtained by checking the band dispersions as shown in Fig.~\ref{fig:fig2_bands} (a).}
The two bands near the Fermi energy with the inverted structure along the $\Gamma$-$\rm M$ ($k_x$) and $\Gamma$-$K$ ($k_y$) directions lead to $c_0>0$ and $c_{1,2}<0$.
While the energy dispersions in the $\Gamma$-$A$ ($k_z$) direction is in normal order, which indicates that $c_3>0$.
Because the two considered bands ($d_{xy}$ and $p_z$) have opposite mirror eigenvalues for the $k_z=0$ mirror reflection, $M_z$ may be presented by $M_z=\sigma_z$.
The mirror reflection symmetry indicates that
\begin{equation}
M_z H_0^{\Gamma}(k_x,k_y,k_z) M_z^{-1}=H_0^{\Gamma}(k_x,k_y,-k_z),\label{eq:MzHMz}
\end{equation}
which leads to
\begin{equation}
d_{x,y}(k_x,k_y,k_z)=-d_{x,y}(k_x,k_y,-k_z),\label{eq:Mz_dxy}
\end{equation}
\begin{equation}
d_{z}(k_x,k_y,k_z)=d_{z}(k_x,k_y,-k_z),\label{eq:Mz_dz}
\end{equation}
For the mirror symmetry in the $k_y=0$ plane, we may write the mirror operator $M_y=-\sigma_z$.
This symmetry places additional constraints to $d_i$,
\begin{equation}
d_{x,y}(k_x,k_y,k_z)=-d_{x,y}(k_x,-k_y,k_z),\label{eq:My_dxy}
\end{equation}
\begin{equation}
d_{z}(k_x,k_y,k_z)=d_{z}(k_x,-k_y,k_z).\label{eq:My_dz}
\end{equation}

Equations~\ref{eq:Mz_dxy} and \ref{eq:Mz_dz} indicate that on the plane $k_z=0$, $d_{x,y}(k_x,k_y,k_z)$  vanish. The solutions of $d_z(k_x,k_y,0)=0$ determine the bands crossing points on the $k_z=0$ plane. Referring to the formula of $d_z$ shown in Eq.~\ref{eq:dz}, we have $c_0+c_1 k_x^2+c_2 k_y^2=0$, with $c_0>0$ and $c_{1,2}<0$, which leads the band-crossing points to form a circle in the $k_x$-$k_y$ plane, namely the nodal ring circling the $\Gamma$ point as shown in Fig.~\ref{fig:fig3_soc} (b).
In the $k_y=0$ plane, Eqs.~\ref{eq:My_dxy} and \ref{eq:My_dz} show that $d_{x,y}$ vanish and the equation
$d_z(k_x,0,k_z)=0$ leads to $c_0+c_1 k_x^2+c_3 k_z^2=0$. Because $c_0,c_3>0$ and $c_1<0$, the solutions
form hyperbolic lines crossing the $k_y$ axis and embracing the $\rm M$, not a circle surrounding the $\Gamma$ point.
These node lines are part of the nodal ring around the $\rm M$ point that will be discussed later.
The last two symmetry operators, $C_3$ and $C_2$ rotations, which place additional symmetrical constraints on $d_i(k_x,k_y,k_z)$, only change
the shape of the nodal rings but not affect its existence.
% In a short summery, the mirror symmetry in the $k_z$ plane and the bands inversion in the $k_x$ and $k_y$
% direction protect the nodal ring $A$ as shown in Fig.~\ref{fig:fig3_nodal_chain}.

The existence of a nodal ring in the $k_y=0$ plane encircling the $\rm M$ point can be proved with a similar argument.
We give the details below. The little group at the $\rm M$ point is $C_{2v}$. It includes the $C_2$ rotation along $x$ axis,
the mirror reflection $M_y$: $(x,y,z)\rightarrow (x,-y,z)$,
and the mirror reflection $M_z$: $(x,y,z)\rightarrow (x,y,-z)$.
The general two bands model around the $\rm M$ point is given as
\begin{equation}
H^M_0({\bf k})=\sum_{i=x,y,z}{g_i({\bf k})\sigma_i},\label{eq:Hm_kp}
\end{equation}
where $g_i(\bf k)$ is real function and vector ${\bf k}$ is related to
the $\rm M$ point. The time-reversal symmetry leads to $g_{x,z}({\bf k})$ being even and $g_y({\bf k})$ being an odd function of $\bf k$.
Generally, $g_z(\bf k)$ may be written as
\begin{equation}
g_z({\bf k})=a_0+a_1 k_x^2+a_2 k_y^2+a_3 k_z^2,\label{eq:gz}
\end{equation}
up to the second order of $\bf k$.
The bands inversion along $\rm M$-$\Gamma$ ($k_x$) and $\rm M$-$\rm L$ ($k_z$) directions lead to $a_0>0$ and $a_{1,3}<0$.
While in the $\rm M$-$\rm K$ ($k_y$) direction the two bands near Fermi energy are in normal order which indicates $a_2>0$.
The two crossing bands near the $\rm M$ point ($d_{xz}$ and $p_y$) with opposite mirror eigenvalues indicate that the mirror reflection $M_y$ may be presented by $M_y=\sigma_z$, which lead to
\begin{equation}
g_{x,y}(k_x,k_y,k_z)=-g_{x,y}(k_x,-k_y,k_z),\label{eq:My_gxy}
\end{equation}
\begin{equation}
g_{z}(k_x,k_y,k_z)=g_{z}(k_x,-k_y,k_z).\label{eq:My_gz}
\end{equation}
From Eqs.~\ref{eq:My_gxy}, \ref{eq:My_gz}, we know that the solutions of $g_z(k_x,0,k_z)=a_0+a_1 k_x^2+a_3 k_z^2=0$ determine the band crossing positions, which form a close circle circling the $\rm M$ point in the $k_x$-$k_z$ plane, namely the nodal ring as shown in Fig.~\ref{fig:fig3_soc} in green.
The $C_2$ rotations constraint the shape of the nodal rings but keep the existence of it.
The mirror symmetry on the $k_z=0$ plane leads to a part of the nodal line embracing the
$\Gamma$ point as discussed above.
Therefore, we prove that there are two types of nodal rings, the red one and the green one
as shown in Fig.~\ref{fig:fig3_soc}, which touch at some point along the $\Gamma$-$\rm M$ direction and form a nodal chain structure in the whole BZ.

\vspace{1mm}
\noindent\textit{Weyl nodes in HfC. ---}
{Generally, in the presence of SOC,
the nodal line structure has the possibility of decaying into Weyl points~\cite{Weng_WSM_PRX_2015,Huang:2015ic}.
The appearance of Weyl points depends on the symmetries of the crystal
structure, the atomic orbital components of the bands forming the nodal lines and the strength of the spin-orbit interaction.
{We derive the effective SOC Hamiltonian for HfC and show the existence of Weyl points in HfC with proper strength of SOC in the Supplemental Material~\cite{suppMat}.}
}
{In order to search the possible band closing points in HfC, we have generated atomiclike
Wannier functions for Hf 5d and C 2p orbitals using the scheme described in Refs.~\cite{Mostofi:2008ff,weng_revisiting_2009,Marzari:2012eu}. The Wannier function without and with SOC
are given in~\cite{suppMat}.}
In the presence of SOC, the bands structure in Fig.~\ref{fig:fig3_soc} shows that the
nodal points along $\Gamma$-$\rm M$, $\rm M$-$\rm L$ and $\rm M$-$\rm A$ are fully gapped.
By careful checking of  the bands structure away form the high symmetry-lines, we find that there are three types, 30 pairs,
of band-crossing points, namely the Weyl points, in the first BZ.
We have located the positions and energies of the three types of Weyl nodes as listed in Table ~\ref{tab:WPposition}.
All others can be found from the listed one by the time reversal and crystal symmetries as shown in Fig.~\ref{fig:fig3_soc} (b).
{By checking the Berry curvature near these Weyl points, the chirality of each Weyl points can be determined~\cite{suppMat}. The red and blue colors indicate the Weyl point with
chirality $\chi=+1$ and $\chi=-1$, respectively.
}

\begin{table}[h]
\begin{centering}
\begin{tabular}{c | c | c}
\hline
 Weyl point & Position ($\rm \AA^{-1}$ )& Energy (eV)\tabularnewline
\hline
$W_{1}$ & (0.7767,\; 0.0112,\; 0.0657) & -0.098 \tabularnewline
$W_{2}$ & (1.0822,\; 0.0311,\; 0.1952) & 0.029  \tabularnewline
$W_{3}$ & (0.6024,\; 0.3463,\; 0.0127) & 0.103  \tabularnewline
\hline
\end{tabular}
\end{centering}

\caption{The three nonequivalent Weyl points in the ($k_x,k_y,k_z$) coordinates shown in
Fig.~\ref{fig:fig3_soc} (b). }\label{tab:WPposition}
\end{table}

\begin{figure}[h]
\begin{centering}
\includegraphics[width=0.5\columnwidth]{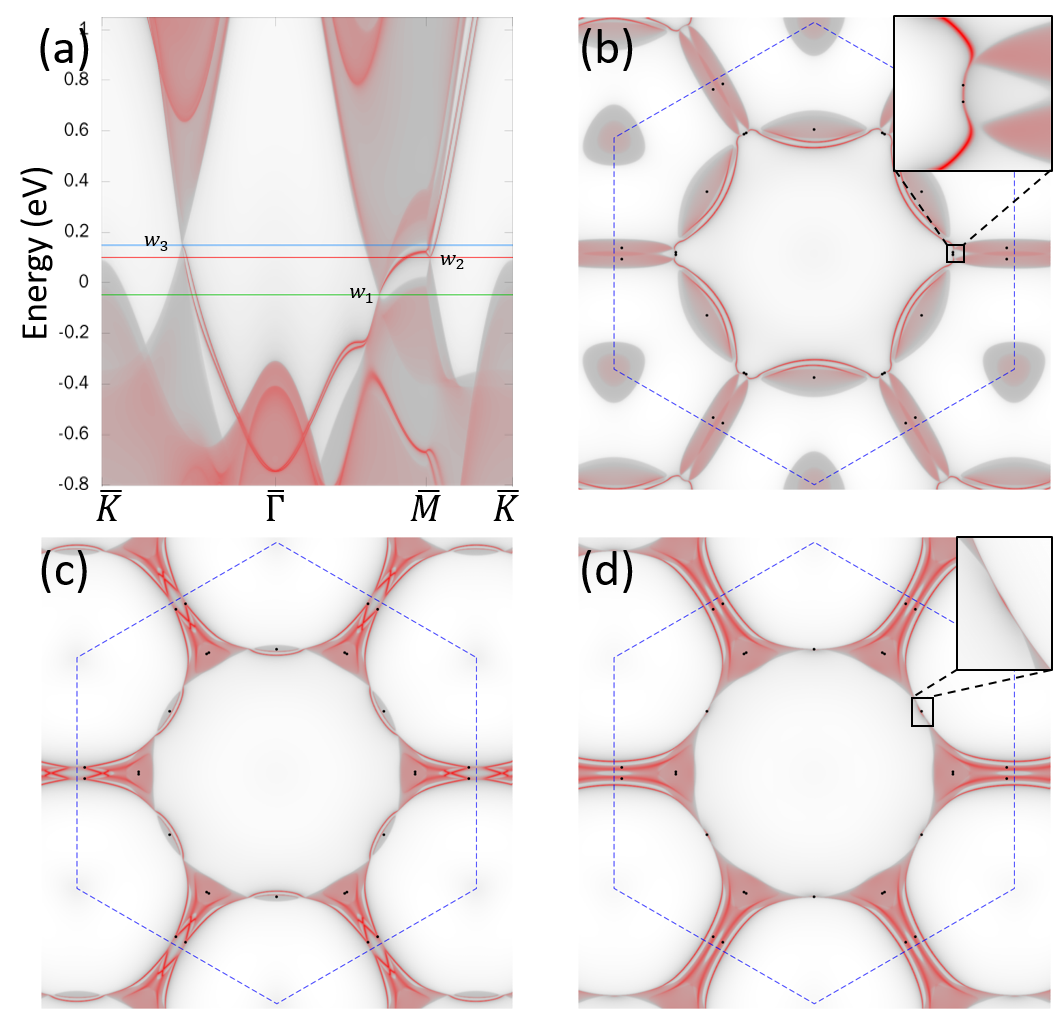}
\end{centering}
\protect\caption{\label{fig:fig4_001}
(Color online) (a) Surface states of HfC terminated in (001) direction.
(b-d) Fermi surface of HfC (001)-surface states. Chemical
potential is set at  (b) $W_1$, (c) $W_2$, and (d) $W_3$ type of Weyl point.
}
\end{figure}

\vspace{1mm}
\noindent\textit{Surface state. ---}
{Based on the tight-binding Hamiltonian generated by the previously obtained Wannier
functions, we have computed the (001) surface states for HfC as shown in Fig.~\ref{fig:fig4_001}.
The projections of Weyl nodes are indicated as black points in Fig.~\ref{fig:fig4_001} (b-d).
Because the Weyl points with left- and right-handed chirality projected
to the same point on the surface BZ,
which does not guarantee the existence of the Fermi arcs,
nor forbid their appearance.
By tuning the chemical potential to cross the three types of Weyl points, it can be seen that
there is one `in' and one `out' Fermi arc connecting each projection of Weyl points
as shown in Figs.~\ref{fig:fig4_001} (b-d).
For instance, if the chemical potential is set at the $\rm W_1$ type of Weyl
points, an extremely long Fermi arc links all the projections of the $\rm W_1$ type of Weyl
points and forms an inward curved hexagon shape
as shown in Fig.~\ref{fig:fig4_001} (b).
The (100) surface states are given in~\cite{suppMat}. }

\vspace{1mm}
\noindent\textit{Conclusions. ---}
% {\section {Conclusions}}
In this Letter we report the nodal chain structure in WC-type HfC without spin-orbit coupling.
Based on first-principles calculation and model analysis we confirm that two types of nodal rings circling
the $\Gamma$ and $\rm M$ points, respectively, connect with each other in the $\Gamma$-$\rm M$ direction
and form a nodal chain structure.
Our finding is  remarkably distinguished from the nodal chain
discussed in Ref.~\cite{nodalchain_Nat_2016}, where nonsymmorphic space group with glide plane symmetry is necessary.
In our Letter, the finding nodal chain is protected by mirror symmetry, which may
happened in simple space groups for a spinless system. This property makes it possible
to design the nodal chain states in weak SOC electronic systems, photonic crystal systems, and phononic crystal systems.
After taking the SOC into consideration, the nodal chain structure degenerates into 30 pairs of Weyl points.
We study the surface states on the (001) and (100) surfaces and show the
patterns for the Fermi arcs connecting the Weyl points with different Fermi energy.

\vspace{1mm}
{\it {Acknowledgments.}}
% \begin{acknowledgments}
{The authors thank Chen Fang and Xi Dai for very helpful discussions.}
This work was supported by the National Key Research and Development Program of China
(No. 2017YFA0304700), the National Natural Science Foundation of China
(No.11422428, No.11674369, No.11404024 and No.11674077), the National Key Research and Development Program of China (No.2016YFA0300600),
the 973 program of China (No.2013CB921700) and the Strategic Priority Research Program (B) of the Chinese Academy of Sciences (No.XDB07020100).
R.Y. acknowledges funding form the National Thousand Young Talents Program.
Q.S. W was supported by Microsoft Research and the Swiss National Science Foundation through the National Competence Centers in Research MARVEL and QSIT.
The numerical calculations in this paper have been done on the supercomputing system in the Supercomputing Center of Wuhan University.
% \end{acknowledgments}

\bibliographystyle{apsrev4-1}
\bibliography{arefs}

\end{document}